\documentclass[%
 reprint,
prstab,
superscriptaddress,
]{revtex4-1}

\usepackage{graphicx}
\usepackage{dcolumn}
\usepackage{bm}
\usepackage{import}
\usepackage[separate-uncertainty=true]{siunitx}
\usepackage{amsmath}
\usepackage{xcolor}
\usepackage{ulem}
\usepackage{subfig}
\usepackage{amssymb}
\usepackage{booktabs}
\usepackage[colorlinks=true,linkcolor=blue,allcolors=blue]{hyperref}%

\begin{document}

\preprint{APS/123-QED}

\title{Modulational instability in large-amplitude linear laser wakefields} 

\author{A. von Boetticher}
\affiliation{Clarendon Laboratory, University of Oxford, Parks Rd, Oxford OX1 3PU, United Kingdom}
\affiliation{Rudolf Peierls Centre for Theoretical Physics, Parks Road, Oxford OX1 3PU, United Kingdom
}
\author{R. Walczak}
\affiliation{Clarendon Laboratory, University of Oxford, Parks Rd, Oxford OX1 3PU, United Kingdom}
\affiliation{Somerville College, Woodstock Road, Oxford OX2 6HD, UK.}
\author{S. M. Hooker}%
\affiliation{Clarendon Laboratory, University of Oxford, Parks Rd, Oxford OX1 3PU, United Kingdom}
\email{simon.hooker@physics.ox.ac.uk}

\date{\today}

\begin{abstract}
We investigate the growth of ion density perturbations in large-amplitude linear laser wakefields via two-dimensional particle-in-cell simulations. Growth rates and wave numbers are found to be consistent with a longitudinal strong-field modulational instability (SFMI). We examine the transverse dependence of the instability for a Gaussian wakefield envelope and show that growth rates and wavenumbers can be maximised off-axis. On-axis growth rates are found to decrease with increasing ion mass or electron temperature. These results are in close agreement with the dispersion relation of a Langmuir wave with energy density that is large compared to the plasma thermal energy density. The implications for wakefield accelerators, in particular multi-pulse schemes, are discussed.

This article was published in \textit{Physical Review E} \textbf{107}, L023201 on 22 February 2023. DOI: \url{10.1103/PhysRevE.107.L02320} \copyright 2023 American Physical Society.
\end{abstract}

\maketitle

Modulational instabilities are a ubiquitous phenomenon, occurring, for instance, in surface water waves \citep{Benjamin:1967}, in optical fibers \citep{Tai_1986} and fusion plasmas \citep[cf.][]{Diamond_2005}. They also arise in plasma-based accelerators \citep[][]{Moulin_1994}, which are capable of generating high-energy particle beams comparable to those of conventional large-scale accelerators over substantially shorter accelerating distances. By using a laser pulse to excite a trailing Langmuir wave in a plasma that acts as an accelerating structure, the laser-wakefield accelerator \citep[LWFA,][]{Tajima_1979} has demonstrated GeV-scale electron acceleration in accelerator stages only a few centimetres long \citep{Leemans_2006}. While such results demonstrate the potential of the technology, laser wakefield accelerators so far rely on large laser systems to generate the required short ($\ll \SI{100}{fs}$), high-intensity ($\sim 10^{18}\SI{}{W cm^{-2}}$) driving pulses. These laser systems have a low wall-plug efficiency ($<0.1\%$) and low pulse repetition rates ($\sim \SI{1}{Hz}$). In contrast, many potential applications of LWFAs --- such as driving compact sources of incoherent X-rays for imaging \citep[][]{Ben-Isma_2011,Kneip_2011,Hussein_2019}, free-electron lasers \citep[][]{Wang_2021}, and, in the longer term, novel particle colliders  --- require efficient laser drivers capable of operating at kHz or higher repetition rates. A possible solution is the multi-pulse laser wakefield accelerator scheme \citep[MP-LWFA, cf.][]{Umstadter_1994, Dalla_1994, Umstadter_1995, Hooker_2014, Cowley_2017}, in which the driving laser energy is delivered over many plasma periods by a train of lower energy laser pulses, spaced by the plasma period. Recent work has identified a route to achieving GeV-scale, kHz repetition rate MP-LWFAs driven by existing laser technology \citep{Jacobsson_2021}.

In laser wakefield accelerators, the dynamics of the plasma ions are not important when the Langmuir wave is excited by a single, high-intensity pulse, owing to the small electron-to-ion mass ratio. Ion motion has been shown to be significant for schemes that use long pulses such as beat-wave accelerators \citep[][]{Mora_1988} or in beam driven devices \citep[cf.][]{Rosenzweig_2005,Vieira_2012,Benedetti_2017}. Ion motion can similarly become important in MP-LWFAs if the plasma wave is excited by a sufficiently long train of individual laser pulses \citep{Hooker_2014}.

The interaction of Langmuir waves with the ion species of a plasma was first investigated in the context of weak plasma turbulence, leading to the discovery of the modulational instability that acts as a dissipation mechanism for wave energy \citep{Vedenov_1964}. If the energy of the Langmuir wave is small relative to the plasma thermal energy, $W := \epsilon_0 E^2 / (2 n_e k_{\mathrm{B}} T_e) \ll 1$, where $E$ denotes the Langmuir wave electric field strength, the modulational instability is described by the Zakharov equations \citep{Zakharov_1972}. These model the non-linear interaction of the fast electrostatic Langmuir oscillation with slow ion motion: small perturbations of the ion density modulate the Langmuir wave envelope; the resulting ponderomotive force associated with the non-uniform Langmuir envelope can in turn reinforce the ion perturbations. The ensuing instability leads to the breakup of the Langmuir wave envelope and eventual Landau damping of short-wavelength modes. In wakefield accelerators, typically $W \gg 1$, precluding the applicability of the Zakharov model. Investigations of this strong-field regime using small mass-ratio approximations by \citet{Silin_1965, Silin_1967} and \citet{Sanmartin_1970} showed that the modulational instability also arises for large-amplitude Langmuir waves. Indeed the development of the modulational instability in plasma wakefields has been inferred in beat-wave and self-modulated laser wakefield experiments \citep{Moulin_1994, LeBlanc_1996, Ting_1996, Marques_1996, Dyson_1996, Kotaki_2002}, but few investigations exist that combine numerical simulation with theory and, thus far, these have been restricted to one spatial dimension \citep{Mora_1988,Heron_1993}.

Studies of the modulational instability in plasma are therefore of both fundamental interest and important in determining the performance, and potential operating regimes, of laser- and beam-drive plasma accelerators. In this Letter we use two-dimensional particle-in-cell (PIC) simulations to investigate the modulational instability in large amplitude, linear laser wakefields. The dominant wave number of the observed perturbations of the ion density, the rate of growth of this mode, and the transverse variations of these quantities, are all found to be in good quantitative agreement with a one-dimensional, analytic theory of the modulational stability in the strong-field regime developed by Silin \citep{Silin_1967} and Sanmartin \citep{Sanmartin_1970}, when the transverse variation of the wakefield amplitude is accounted for. The dependence on ion mass and electron temperature of the growth rate of the ion density perturbations observed in the PIC simulations are also found to be in good agreement with the analytic theory, providing further evidence that the perturbations arise from the modulational instability. The simulations performed in this work are, to the best of our knowledge, the first multi-dimensional PIC simulations of the modulational instability in a laser wakefield and the first to test the Silin-Sanmartin theory in the high-amplitude regime.

PIC simulations of laser wakefields were performed with the relativistic PIC code \texttt{smilei} \citep[Simulation of Matter Irradiated by Light at Extreme Intensities,][]{SMILEI}. The simulations used ${N}_{\mathrm{PPC}} = 64$ particles per cell, a longitudinal grid spacing, $\Delta x$, equal to the Debye length of the plasma, $\lambda_D$, and transverse grid spacing $\Delta y = 5 \lambda_D$. The simulations were performed in a two-dimensional geometry containing an electron-proton plasma slab with longitudinal and transverse extent of 280 $\mu$m and 160 $\mu$m, respectively, at a uniform plasma density, $n_e = 9.7\times 10^{17}$ cm$^{-3}$. The electron temperature was initialized near the ionisation energy of hydrogen, $T_e = 15$ eV$k_\mathrm{B}^{-1}$; the initial ion temperature was set to $T_i = 0.075$ eV$k_\mathrm{B}^{-1}$, thus $T_i \ll T_e$ as is typical of a plasma ionised by a short laser pulse. 
A high-intensity laser pulse with Gaussian temporal and transverse spatial envelope was propagated through the plasma, generating a periodic oscillation of the electron density at the plasma frequency, $\omega_{pe} = 5.59 \times 10^{13}\ \mathrm{rad\ s}^{-1}$. The full width at half-maximum (FWHM) pulse duration and width were set to $\tau_{\mathrm{fwhm}} = 45$ fs and $w_{\mathrm{fwhm}} = 45$ $\mu$m, respectively, with a central laser wavelength of $\lambda_l = 802.5$ nm and peak intensity $I_{\mathrm{peak}} = 5.3 \times10^{17}$ Wcm$^{-2}$, corresponding to a normalised vector potential, $a_0 = 0.5$. These parameters were selected to obtain a wakefield with a density amplitude variation $\delta n_e / n_{e}$ on the order of $10\%$, corresponding to a regime with large wakefield amplitude but no strong nonlinear relativistic effects. The simulation properties and laser and plasma parameters are summarised in Table 1 in the Supplemental Material \citep[][]{Supp_mat} to this publication, which includes references \citep{Lifshitz_2009,Thornhill_1978,Brent_1973,scipy}.
\begin{figure}[t]
\centering
\includegraphics[width=0.8\columnwidth]{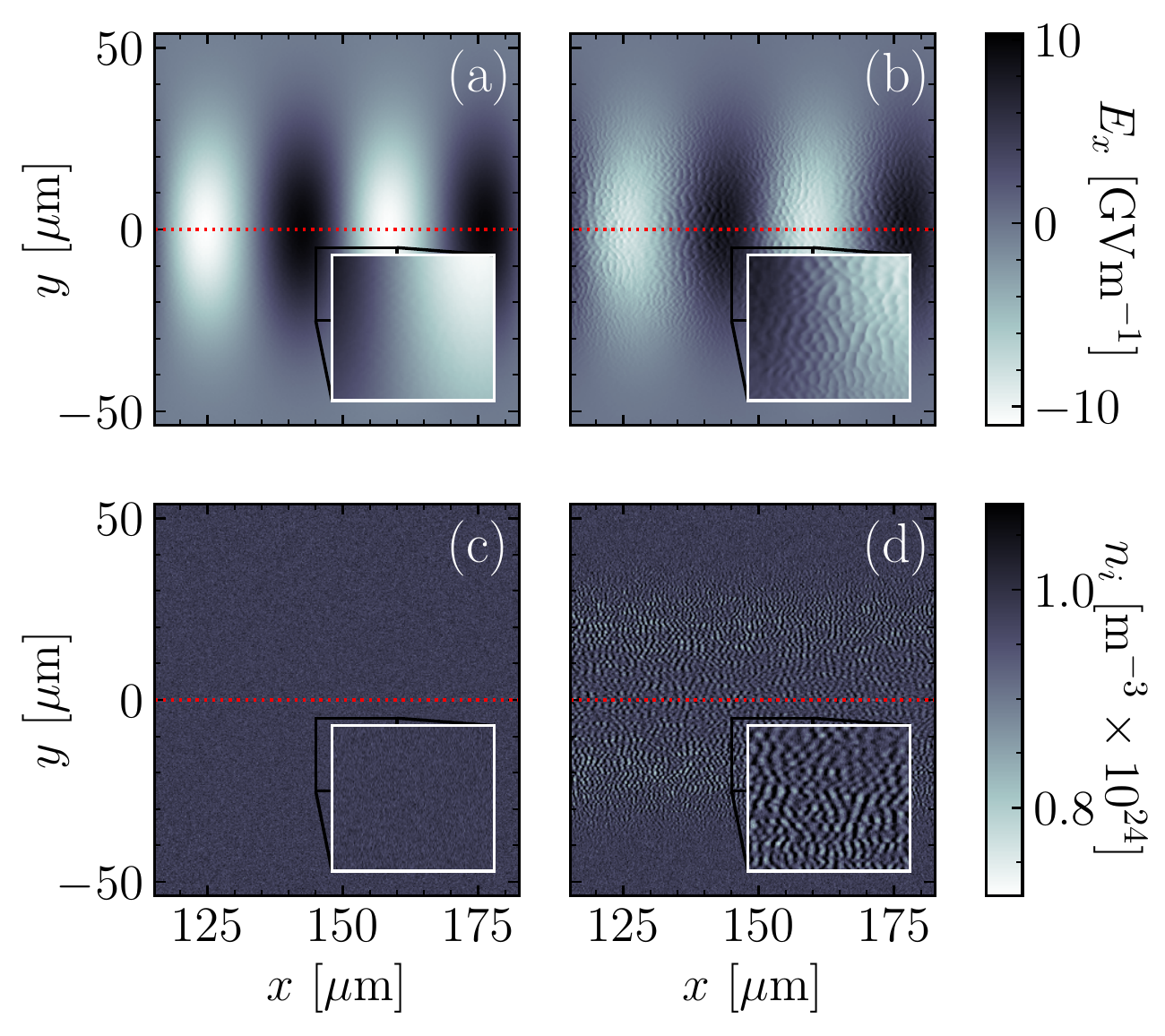}
\caption{(a-b) Longitudinal electric field strength $E_x(x,y)$ as a function of longitudinal ($x$) and transverse ($y$) position, (a) 0.32 ps and (b) 2.56 ps after wake excitation, showing the development of high-frequency modulations of the field envelope, magnified in the inset. (c-d) Corresponding ion density $n_i(x,y)$, showing the development of high-frequency modulations. The red line indicates the axis of initial laser propagation.} \label{field_modulations}
\end{figure}

The wakefield excited by the laser pulse attains a maximum longitudinal electric field strength $E_x = 8.5$ GVm$^{-1}$ on axis, wavelength $\lambda_{\mathrm{wake}} = \lambda_{pe} \approx 2\pi c / \omega_{pe} = 33.7$ $\mu$m and oscillates at the plasma frequency, since $\omega_{k,pe}^2 = \omega_{pe}^2 + 3k^2v_{th,e}^2 \approx \omega_{pe}^2$ for wavelengths $\lambda = 2\pi/k$ that are long compared to the Debye length, $\lambda_D = (\epsilon_0 k_\mathrm{B} T_e / n_e e^2)^{1/2} = 29.1$ nm. Here $v_{th,e} = (k_B T_e / m_e)^{1/2}$ denotes the electron thermal speed.
The longitudinal electric field of the wakefield is shown in Fig. \ref{field_modulations}(a), at $t = 0.32$ ps after the wakefield excitation. The on-axis longitudinal electric field strength is associated with a sinusoidal plasma density oscillation observed in the simulations of amplitude $\delta n_e / n_e = 0.09$. The transverse profile of the longitudinal electric field envelope is consistent with the laser pulse profile, with a transverse width $w_{\mathrm{fwhm}} = 45$ $\mu$m.

\begin{figure}[t]
    \captionsetup[subfloat]{justification=centering}
    \centering
    \includegraphics[width=0.8\columnwidth]{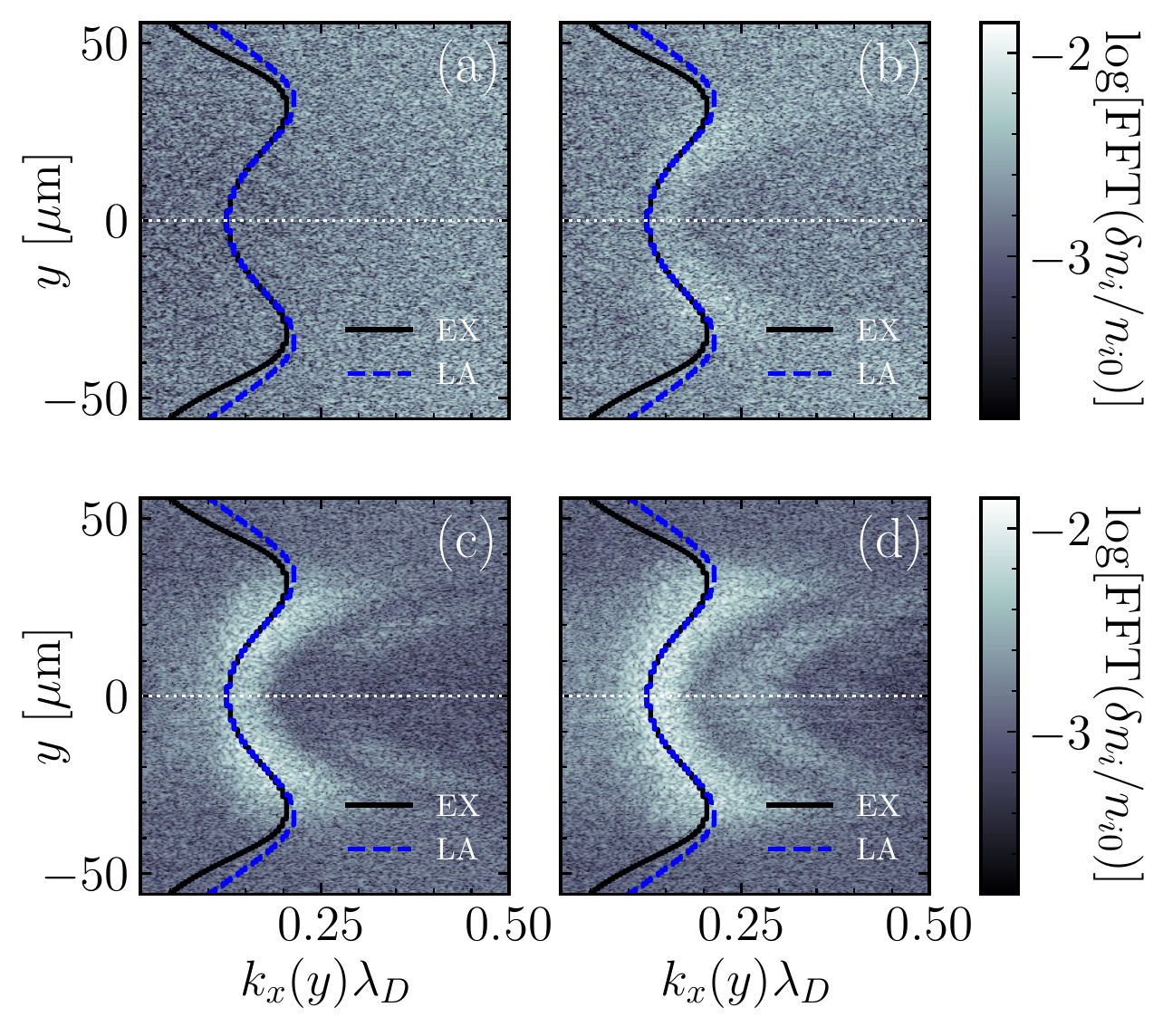}
    \caption{Longitudinal wave number spectrum $\log_{10}[\mathrm{FFT}_x(\delta n_i(x,y)/n_{i0})]$ of the perturbed ion density, as a function of transverse position $y$, at (a) $t = 0.64$ ps after wake excitation, (b) $t=1.60$ ps, (c) $t=2.56$ ps and (d) $t = 3.52$ ps. The instability initially grows fastest off-axis (b). The predicted transverse wave number dependence of the fastest growing mode is calculated by solving the exact Eq. \eqref{Sanmartin dispersion relation} (EX, black, solid line) and the approximate Eq. \eqref{Sanmartin approximate dispersion} valid for a large driving field amplitude (LA, blue, dashed).} \label{Ion density FFT}
\end{figure}

Within two picoseconds of the wakefield excitation, modulations of the longitudinal electric field and the ion density are observed to develop, with a characteristic length of $1.4$ $\mu$m on-axis ($y = 0$) that is much smaller than the wakefield wavelength, $\lambda_{pe} = 33.7$ $\mu$m. The modulations of the field and ion density are visible in Fig. \ref{field_modulations}(b) and \ref{field_modulations}(d), respectively, at $t = 2.56$ ps after wakefield excitation, and are shown magnified in the insets. The timescale for the growth of the modulations is thus on the order of the ion plasma period, $2\pi\omega_{pi}^{-1} = 4.8$ ps. The wave number spectrum of the ion density is shown in Fig. \ref{Ion density FFT} as a function of transverse position, indicating that modulations of the ion density grow fastest away from the axis, and that the wavelength of these modulations decreases with distance from the axis $y=0$. The on-axis spectrum exhibits a peak at a wave number $k \lambda_D \approx 0.13$; a weaker, higher-order mode is seen to develop at $k \lambda_D \approx 0.26$. The modulations of the ion density are phenomenologically consistent with a modulational instability \citep{Vedenov_1964,Zakharov_1972}. 

For the plasma and wakefield parameters encountered here, the ratio of the Langmuir energy to the plasma thermal energy density is $W \approx 150 \gg 1$, precluding the applicability of the Zakharov equations to describe the modulational instability. The theory of the modulational instability in the strong-field regime, $W \gg 1$, was developed by \citet{Silin_1965,Silin_1967} and \citet{Sanmartin_1970}, and is reviewed briefly here and in the Supplemental Material to this paper \citep[][]{Supp_mat}. The dispersion relation for a high-frequency electrostatic oscillation in a Maxwellian plasma obtained by \citet{Sanmartin_1970} simplifies considerably if the frequency of the equilibrium oscillation is large compared to the ion plasma frequency, $\omega_0 \gg \omega_{pi}$, and wavelengths of all modes are long compared to the electron Debye length, $k \lambda_D = k v_{th,e}/\omega_{pe} \ll 1$, as is the case for the wakefield oscillation and ion density modulations observed in the PIC simulations. Since the wakefield is in resonance with the plasma frequency, the resonance parameter $\tilde{\lambda} := (\omega_0^2 - \omega_{k,pe}^2)/\omega_{pe}^2 \ll 1$ is small. With $\mu := m_e / m_i \ll 1$, and assuming a purely growing mode with growth-rate $\gamma \in \mathbb{R}$, the dispersion relation is given by \citep{Sanmartin_1970}
\begin{align}
    1 + \chi_i&\bigg[1 - 2\frac{J_0(a)J_1(a)}{a} + k^2\lambda_D^2 \nonumber \\ &-2J_1(a)^2\frac{\tilde{\lambda}-\frac{1}{4}\tilde{\nu}^2 + 2\mu\left[1-J_0(a)^2\right]a^{-2} }{\tilde{\lambda}^2+(2\tilde{\gamma}+\tilde{\nu})^2}\bigg] = 0, \label{Sanmartin dispersion relation}
\end{align}
where $\gamma$ and $\tilde{\gamma} = \gamma / \omega_{pe}$ denote the growth rate and normalised growth rate, respectively; $\tilde{\nu} = \nu / \omega_{pe} \approx (\pi/2)^{1/2} (k\lambda_D)^{-3}\exp(-1/[2k^2\lambda_D^2])$ is the normalised Landau damping rate, and $J_0$ and $J_1$ denote Bessel functions of the first kind. The parameter $a := e k E_0/m_e\omega_0^2$ measures the ratio of the electron excursion distance in a wakefield cycle to the mode wavelength $2\pi k^{-1}$. The ion susceptibility, $\chi_i(\gamma,k)$, can be related to the plasma dispersion function, $Z$, \citep[cf.][]{Fried_1961}, via
\begin{align}
   \chi_i(\gamma,k)
   &= \frac{\omega_{pi}^2}{k^2} \int_{C_L} \frac{k(dF_{i0}/dv)}{i\gamma - kv}\ dv \\
   &= \frac{\omega_{pi}^2}{k^2v_{th,i}^2}\left[1 + \frac{\zeta_i}{\sqrt{2}} Z(\zeta_i/\sqrt{2})\right] \label{chi}
\end{align}
where $C_L$ denotes the usual Landau contour, $F_{0i} = 1 / (\sqrt{2\pi}v_{th,i}) e^{-v^2/2v_{th,i}^2}$, $\zeta_i := i\gamma/kv_{th,i}$, and the plasma dispersion function is given by
\begin{align}
    Z(\zeta_i) := \frac{1}{\sqrt{\pi}} \int \frac{e^{-u^2}}{u - \zeta_i}\ du.
\end{align}
For calculation it is useful to relate $Z$ to the error function via $Z_i(\zeta_i) = i \sqrt{\pi} \exp(-\zeta_i^2)\left [1 +  \mathrm{erf}(i\zeta_i)\right]$ \citep[][]{Fried_1961} so that,
with $z := -i\zeta_i = \gamma/kv_{th,i}$, the ion susceptibility can be expressed as
\begin{align}
    \chi_i(z,k) = \frac{\omega_{pi}^2}{k^2v_{th,i}^2}\bigg[1 &- z\sqrt{\frac{\pi}{2}} \exp\left(\frac{z^2}{2}\right)\nonumber \\&\times \left(1 - \mathrm{erf}\left[z/\sqrt{2}\right]\right)\bigg]. \label{comp}
\end{align}
This expression can be evaluated using standard implementations of the error function. 

\begin{figure}[t!]
    \raggedright
    \hspace{0.5cm}
    \includegraphics[width=0.75\columnwidth]{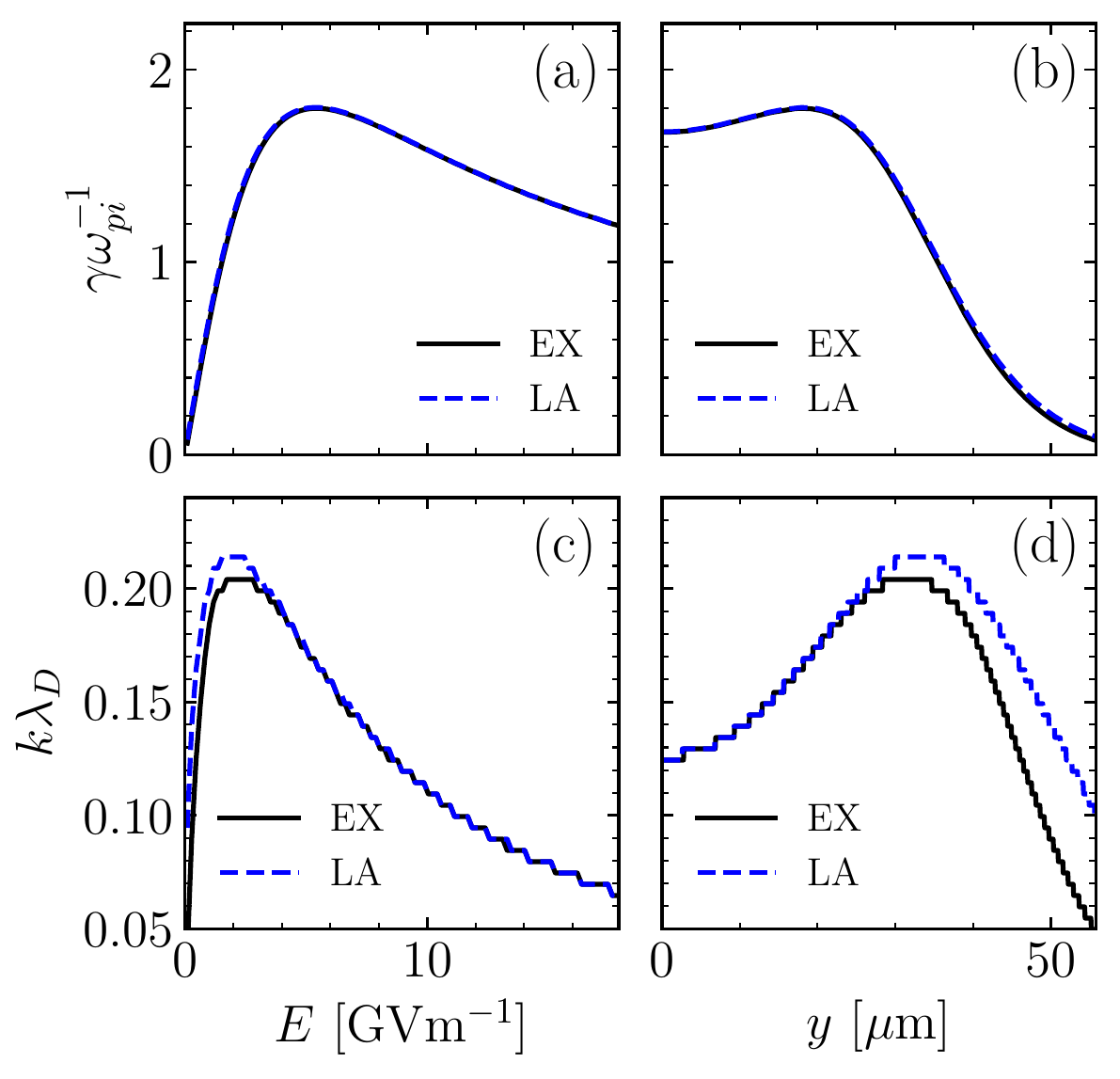}
    \caption{Growth rates and wave numbers calculated with the exact (EX) Eq. \eqref{Sanmartin dispersion relation} and the large-amplitude approximation (LA) Eq. \eqref{Sanmartin approximate dispersion}. (a) Variation of the maximum growth rate with the wakefield electric field strength, (b) transverse dependence of the growth rate, for a transverse Gaussian wakefield amplitude profile, with $E_x(y=0) = 8.5$ GVm$^{-1}$ and $w_{\mathrm{fwhm}} = 45$ $\mu$m, (c) variation of the dominant wave number with electric field strength, and (d) transverse dependence of the wave number for a transverse Gaussian wakefield profile.}
    \label{growth rate dependencies}
\end{figure}

Eq. \ref{Sanmartin dispersion relation} can be simplified for sufficiently fast growth rates. The assumption $k\lambda_D \ll 1$ together with $T_i / T_e \ll 1 $ implies that $kv_{th,i} = k \lambda_{D} \sqrt{T_i/T_e} \omega_{pi} \ll \omega_{pi}$. For the growth time scales observed in the PIC simulations, $\gamma^{-1} \sim \omega_{pi}^{-1}$, it follows that, indeed, $z \gg 1$. The ion susceptibility can then be approximated using the asymptotic expansion of the plasma dispersion function $Z(\zeta) = -\zeta^{-1} - \zeta^{-3}/2 + O(\zeta^{-5})$ \citep{Fried_1961}. Using this in Eq. \ref{chi}, one obtains $\chi_i \approx \omega_{pi}^2/\gamma^2$. The dispersion relation, Eq. \ref{Sanmartin dispersion relation}, then simplifies to the quartic \citep[][]{Sanmartin_1970}
\begin{align}
    &\left[\tilde{\lambda}^2 + (2\tilde{\gamma} + \tilde{\nu})^2\right]\left[\tilde{\gamma}^2 + \mu\left(1-\frac{2}{a}J_0(a)J_1(a)\right)\right] \nonumber \\
    &- 2\mu J_1^2(a)\left[\tilde{\lambda} - \frac{1}{4}\tilde{\nu}^2 + 2\mu \left[1-J_0(a)^2\right]a^{-2}\right] = 0.\label{Sanmartin approximate dispersion}
\end{align}
We numerically solved Eq. \eqref{Sanmartin approximate dispersion} and examined the root with $\max\{\gamma > 0\}$, corresponding to the fastest growing instability. The dependence of the maximum growth rate and the associated dominant wave number on the wakefield electric field strength is shown in Fig. \ref{growth rate dependencies}(a) and \ref{growth rate dependencies}(c), respectively. The growth rate of the modulational instability initially increases with the field strength, is maximised for $E_0 \approx 5.5$ GVm$^{-1}$ and decreases for stronger fields. This non-monotonic behaviour of the growth-rate is associated with the electron displacement in the wakefield oscillations. When the electron excursion distance significantly exceeds the mode wavelength, corresponding to $a \gg 1$, the coupling between the wakefield and the ion perturbations becomes weak as the electrons experience a spatially averaged pondermotive force \citep{Mora_1988}; indeed for sufficiently strong fields and large electron displacements the modulational instability is predicted to be suppressed entirely \citep{Sanmartin_1970}. 
By assuming a transverse Gaussian profile of the wakefield electric field envelope, $E_x(y) = E_0\exp(-y^2/2\sigma^2)$, with the parameters $E_0$ and $\sigma$ determined from a fit of this expression to the transverse variation of the initial wake amplitude obtained from the PIC simulations, the expected transverse scaling of the dominant wave number can be calculated. To do so, the dispersion relation, Eq. \eqref{Sanmartin approximate dispersion}, was solved on a transverse grid, for the wave number, $k(y)$, that maximises the growth rate. The resulting transverse growth rate and wave number dependence are shown in Fig. \ref{growth rate dependencies}(b) and \ref{growth rate dependencies}(d), respectively. 
For transverse displacements $|y| \lesssim 30$ $\mu$m, the predicted wave number dependence is in good agreement with the transverse variation of the wave number of ion density modulations observed in the PIC simulations, shown in Fig. \ref{Ion density FFT}. For transverse displacements $|y| \gtrsim 30$ $\mu$m, Eq. \eqref{Sanmartin approximate dispersion} predicts a decrease in the wave number with greater transverse displacement; however at these displacements the observed growth of the instability is very slow and is suppressed entirely for $|y| \gtrsim 40$ $\mu$m.
Fig. \ref{Ion density FFT}(b) shows that modulations of the ion density first develop at a transverse displacement $y_{\max\{\gamma\}} \approx \pm 20$ $\mu$m, indicating that the growth rate is maximised off-axis, where the electric field strength of the wakefield is lower than on axis. This is consistent with the simple model of the transverse variation of the growth rate which is shown in Fig. \ref{growth rate dependencies}(b). Fig. \ref{growth rate dependencies}(b) indicates that for the wakefield amplitude and transverse profile considered here the maximum growth rate is indeed expected at $y \approx 20$ $\mu$m.

\begin{figure}[t!]
    \captionsetup[subfloat]{justification=centering}
    \centering
    \includegraphics[width=0.85\columnwidth]{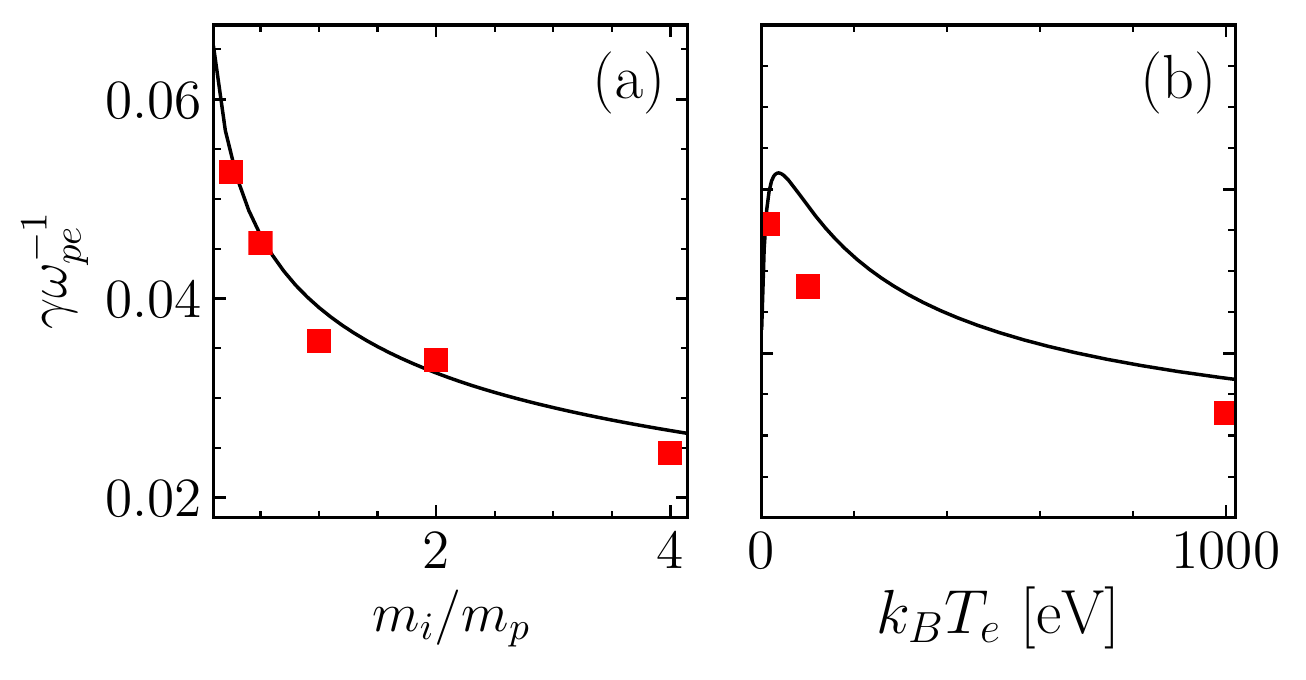}
    \caption{(a) Dependence of the modulational instability growth rate on the ion mass, for $T_e = 15$ eV$k_{\mathrm{B}}^{-1}$, obtained from Eq. \eqref{Sanmartin dispersion relation} (black line) and  PIC simulations (red squares, for $m_i = \{0.25, 0.5, 1.0, 2.0, 4.0 \} m_p$). (b) Dependence of the growth rate on the electron temperature (black line) and measured growth rates from simulations (red squares), for $m_i = m_p$.}\label{mass and temperature dependence}
\end{figure}

To provide further evidence that the modulational instability is responsible for the behaviour observed in Fig. \ref{Ion density FFT}, PIC simulations were performed for different ion masses and electron temperatures, with  other simulation parameters held fixed. Growth rates of the ion density perturbations were determined from the PIC simulations by measuring the amplitude of the fastest growing mode over time \citep{Supp_mat}. Figure \ref{mass and temperature dependence}(a) compares the mass dependence of the growth rates determined this way with that calculated from Eq. \eqref{Sanmartin approximate dispersion}. The two calculations are seen to be in excellent agreement.
Simulations at elevated electron temperatures of $T_e = 100$ eV$k_\mathrm{B}^{-1}$ and $T_e = 1$ keV$k_\mathrm{B}^{-1}$ were performed and growth rates of the ion density perturbations on axis ($y=0$) were compared to the theoretical predictions, shown in Fig. \ref{mass and temperature dependence}(b). The growth rates follow a similar trend with $T_e$, and agree within 30\% for the temperature range investigated, although the rates calculated by the PIC simulations are consistently lower than those deduced from Eq. \eqref{Sanmartin approximate dispersion}. We hypothesise that dimensionality effects may be responsible for the observed differences in measured and predicted growth rates as a function of the electron temperature. The better spatial resolution of the PIC simulations, relative to the Debye length, obtained at higher temperatures may also play a role. The analytic theory predicts a weak resonance in the growth rate near $T_e \approx 40$ eV$k_\mathrm{B}^{-1}$, but there is insufficient data to state whether or not this resonance is also predicted by the PIC simulations. The good quantitative agreement seen in Fig.\ \ref{mass and temperature dependence} with the analytic theory provides further confirmation that the ion density perturbations observed in the PIC simulations arise from the modulational instability. 

The PIC simulations presented here show the development of the strong-field modulational instability in high-amplitude linear laser wakefields. The growth rates, wave numbers, and the transverse profiles thereof are found to be in good agreement with those obtained from the one-dimensional dispersion relation of \citet{Sanmartin_1970} if the transverse dependence of the wakefield amplitude is accounted for. We note that \citet{Vieira_2012} have investigated the role of ion dynamics in self-modulated proton beam driven plasma wakefield accelerators. Those accelerators operate in the narrow beam limit \citep{Vieira_2012}, where the transverse extent of the proton driver, $\sigma_r$, is much smaller than the wavelength of the electron plasma oscillation. In that regime, ion motion is dominated by a strong transverse component, since the ponderomotive force of the wakefield envelope scales with the gradient of the wakefield intensity, $F_{p,\perp} \sim \nabla_\perp E^2 \sim E^2 / \sigma_r$. In contrast, the wakefields investigated here have a transverse scale length set by the laser beam width of $w_{\mathrm{fwhm}} = 45$ $\mu$m (FWHM) that is somewhat greater than the electron plasma wavelength $\lambda_{pe} = 33.7$ $\mu$m, such that longitudinal and transverse effects are comparable. Indeed, the small-scale transverse variations of the modulations visible in Fig. \ref{field_modulations} show that transverse effects are non-negligible.

We now consider the implications of these results for multi-pulse laser wakefield accelerators. Since the modulational instability develops on a timescale comparable to the ion plasma period, it will become relevant for pulse trains with $N_{\mathrm{pulse}} \gtrsim (m_i/m_e)^{1/2}/Z$, where $Z$ denotes the ion charge number. Indeed, in earlier work we showed that for long pulse trains with $N_{\mathrm{pulse}} \gg (m_i/m_e)^{1/2}/Z$, the amplitude of the wakefield did not grow monotonically with pulse number \citep{Hooker_2014}, indicating a loss of coherence of the electron oscillations. This earlier finding is explained by the work reported here. The non-monotonic dependence of the SFMI growth rate on the electric field strength observed transversely in Fig. \ref{growth rate dependencies}(b) suggests that, in principle, the intensities of pulses in the driving train could be tuned to minimise the time for which the wakefield amplitude is such that the SFMI growth-rate is large. Alternatively, using a species with a high $m_i / Z$ ratio, or heating the electronic component of the plasma could reduce the onset of the modulational instability, and allow longer pulse trains to be employed.

Although the effects of the modulational instability could be mitigated by these methods, they are unlikely to be necessary for MPLWFAs. For example,  \citet{Jacobsson_2021} recently introduced a novel method for driving GeV-scale MPLWFAs with a single picosecond-duration pulse by spectrally-modulating the pulse in the wakefield driven by a low-energy seed pulse, and compressing it to generate a train of short pulses. In that work the generated pulse trains have $N_{\mathrm{pulse}} \sim 10 < (m_i/m_e)^{1/2}$. Separate PIC simulations were performed to confirm that a wakefield excited by such short pulse trains is indeed unaffected by ion dynamics during the excitation.

This work was supported by the UK Engineering and
Physical Sciences Research Council (EPSRC) [grant nos EP/N509711/1, EP/V006797/1], the UK Science and Technologies Facilities Council [grant nos. ST/N5044233/1, ST/P002048/1, ST/V001655/1], the European Union's Horizon 2020 research and innovation programme [grant no. 653782], the plasma HEC Consortium [grant no. EP/R029149/1] and the ARCHER UK national supercomputing service. This material is based upon work supported by the Air Force Office of Scientific Research under award number FA9550-18-1-7005.

%Bibliography
%

\end{document}